\documentclass{sig-alternate}

\usepackage{amssymb}
\usepackage{graphicx}
\usepackage{algorithm} 
\usepackage{wrapfig} 
\usepackage{algorithmic} 
\usepackage{breqn}
\usepackage{tabularx} 
\usepackage{subfig} 

\usepackage{multirow}
\usepackage{enumitem}

\usepackage{subfig}
\usepackage{pgf}

\newtheorem{definition}{Definition}
\usepackage{booktabs}

\begin{document}


\title{Learning to Match for Multi-criteria Document Relevance}

\numberofauthors{3}

\author{
\alignauthor Bilel Moulahi\\
      \affaddr{IRIT, Université Paul Sabatier}\\
       \affaddr{118 Route Narbonne, Toulouse, France}\\
     \affaddr{Universit\'e de Tunis El Manar}\\
      \affaddr{Facult\'e des Sciences de Tunis}\\
       \affaddr{LIPAH, 2092, Tunis, Tunisie}\\
       \email{bilel.moulahi@irit.fr}
\alignauthor Lynda Tamine \\
       \affaddr{IRIT, Université Paul Sabatier}\\
       \affaddr{118 Route Narbonne}\\
       \affaddr{Toulouse, France}\\
       \email{tamine@irit.fr}
\alignauthor Sadok Ben Yahia\\
     \affaddr{Universit\'e de Tunis El Manar}\\
      \affaddr{Facult\'e des Sciences de Tunis}\\
       \affaddr{LIPAH, 2092, Tunis, Tunisie}\\
       \email{sadok.benyahia@fst.rnu.tn}
}


\maketitle

\begin{abstract}

In light of the tremendous amount of data produced by social media, a large body of research have revisited the relevance estimation of the users' generated content. Most of the studies have stressed the multidimensional nature of relevance and proved the effectiveness of combining the different criteria that it embodies. Traditional relevance estimates combination methods are often based on linear combination schemes. However, despite being effective, those aggregation mechanisms are not effective in real-life applications since they heavily rely on the non-realistic independence property of the relevance dimensions. In this paper, we propose to tackle this issue through the design of a novel fuzzy-based document ranking model. We also propose an automated methodology to capture the importance of relevance dimensions, as well as information about their interaction. This model, based on the Choquet Integral, allows to optimize the aggregated documents relevance scores using any target information retrieval relevance metric. Experiments within the TREC Microblog task and a social personalized information retrieval task highlighted that our model significantly outperforms a wide range of state-of-the-art aggregation operators, as well as a representative learning to rank methods.
\end{abstract}

\section{Introduction}\label{Introduction}
A large body of research has focused on the core concept of relevance in information retrieval (IR) \cite{borlund2003,saracevic2007III,taylor2012}. While early work particularly considered the topical relevance, others argued that the user context is highly dependent on many relevance factors that represent the basic clue for relevance assessment \cite{saracevic2007III}.
Recently, the increasing availability of user generated content over social media has brought new challenges to multi-relevance estimation because of both the diversity, the task-dependency and the user's context-dependency of the involved relevance features.
For instance, consider a user submitting the query ``\textit{municipal elections}'' in the Twitter search system. In such scenario, the user usually looks for very fresh tweets from her/his region rather than other locations, satisfying as much as possible his/her information need (\textit{i.e.}, topical matching between the query and the tweets). While a typical IR system is able to return recent tweets, it often suffers from coverage and ranking problems, as the user may be interested in more complex ranking scenario encompassing other Twitter quality criteria beyond topicality (\textit{e.g.}, authority, credibility and geo-localization dimensions, etc.).
Interestingly, studies held through \textit{Searchmetrics} \footnote{http://www.searchmetrics.com/en/services/ranking-factors-2013/} highlighted both the diversity of the features used by modern search engines such as Google+, Twitter and Facebook and the important part of social features particularly.

\begin{figure}[htbp]
  \begin{center}
    \includegraphics[width=6.8cm,height=4.8cm]{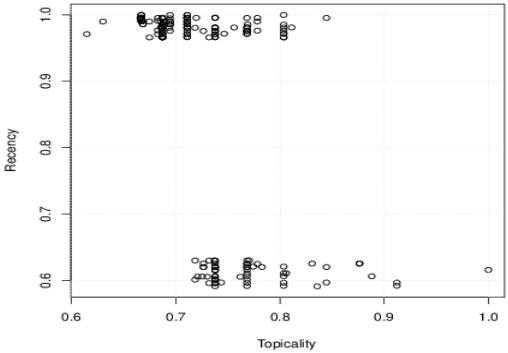}
  \end{center}
  \caption{Distribution of relevance scores for topic $83$, ``\textit{Stuxnet Worm effects}'', of the TREC $2012$ Microblog track.}\label{spearman}
	\end{figure}

A wide range of challenging IR applications including mobile IR \cite{Goker2008}, personalized IR \cite{daoud2010,celia2011,daoud2011,sieg2007} and social IR \cite{metzler2011,duan2010,amjed2012} involve heavily the aggregation of multiple relevance dimensions.
In the sake of addressing this challenge, previous solutions are mostly based on classical aggregation functions such as weighted means or linear combination schemes in the form of products and sums. However, these aggregation operators assume that relevance dimensions are \textit{not independent} of each other and generally those including topical relevance are rated as the highest ones in importance \cite{saracevic2007III,eichkoff2013}.
For instance, we show in Figure \ref{spearman} the distribution of relevance scores of Topic $83$: ``\textit{Stuxnet Worm effects}'' from the TREC\footnote{http://trec.nist.gov} $2012$ Microblog track with respect to recency and topicality criteria. We compute the Spearman Rank correlation coefficient, and we find a notable negative correlation ($\rho=-0.54$), as it may be seen in the lower and higher parts of the graph.
Obviously, the performances scores are considerably high either for the topicality criterion or the recency one.
Considering the case of such correlation scenarios and bearing in mind that classical aggregation operators are assumed to hold the additive property across individual scores, the latter is unable to account for dependencies between those interacting relevance dimensions.
Therefore, using an aggregation mechanism relying on a linear combination scheme can be effective and convenient in some applications, but can also be somewhat inadequate in many IR tasks. Although advanced aggregation operators were recently proposed \cite{celia2011,gerani2012,eichkoff2013}, only a few \cite{eichkoff2013} considered specifically the interactions existing among the relevance dimensions. In practice, this problem is usually avoided by constructing independent criteria \cite{Goker2008}.

In this paper, we assume a more general scenario where different dependent or independent relevance dimensions are considered within a document retrieval task. We propose a new multidimensional relevance aggregation operator based on the discrete Choquet integral \cite{choquet53}. The latter has been successfully applied on a wide variety of domains during the last decades \cite{grabisch2000}. \\
More specifically, the contributions of this paper are twofold and mainly include:
\begin{itemize}
\item A novel multi-criteria aggregation approach to address the problem of combining relevance estimates and a machine learning driven methodology to train the model;
\item A large-scale experimental evaluation within two different IR tasks namely the tweet search task \cite{ounis2011} and the personalized social bookmarking task \cite{vallet2012}.
\end{itemize}
The remainder of the paper is organized as follows. In Section \ref{relworks}, we briefly survey related work to put our contribution in context. Section \ref{choquet} describes our multidimensional relevance aggregation operator. In Sections \ref{experiments} and \ref{evalresult}, we describe the experimental setup and then present the experiments and discuss the obtained results. Section \ref{conclusion} concludes the paper and outlines future work.

\section{Related Work} \label{relworks}
In this section, we briefly overview two research directions related to document relevance estimates combination: (i) the relevance concept and the different features it entails; and (ii) relevance criteria combination mechanisms. We present some approaches that are relevant to multidimensional relevance aggregation, and we reveal their significant differences from standard relevance aggregation mechanisms.

\subsection{On the Diversity of Relevance Features in IR }\label{features}

The relevance concept in IR has gathered a great attention in the two last decades \cite{borlund2003,saracevic2007III}. Although there is no wide consensus, the general relevance level includes many dimensions such as content, object, validity, situational, affective and belief dimensions \cite{saracevic2007III}. Each dimension refers to a group of criteria considered by the users to make relevance inferences \cite{saracevic2007III}.\\
The main findings are threefold:
\begin{itemize}
	\item  Relevance dimensions are \textit{not independent} of each others and generally those related to content, which include topical relevance, are rated as the highest ones in importance, but \textit{interact} with other dimensions \cite{saracevic2007III,eichkoff2013};
	\item A few and finite relevance dimensions are considered jointly by the users to assess relevance;
	\item The importance of the dimensions depends on tasks in progress and class of users.
\end{itemize}
Considering these findings and tasks specificities, works in many recent IR applications such as mobile IR \cite{cong2009,Goker2008}, social IR \cite{metzler2011,duan2010,nagmoti2010} and personalized IR \cite{sieg2007}, attempt to go beyond the classical content-matching dimension. \\
For instance, in mobile IR, relevance is based on the user's situation (location) and social (surrounding persons) dimensions \cite{Goker2008,cong2009}. In social networking services like Twitter, a variety of micro-blogging specific criteria such as topic matching and authority belonging respectively to content and validity dimensions, are investigated to compute relevance \cite{nagmoti2010}.
In the same line of research, authors in \cite{duan2010} propose to rank tweets according to the number of followers, the number of mentions as well as the authority of the microblogger, computed by applying PageRank algorithm on retweet social network. The content relevance feature is always the most important feature for all search settings.
The authors in \cite{amjed2012} proposed new social features for social search in micro-blogging networks such as Twitter. They proposed new relevance factors such as the social importance of microbloggers and the temporal magnitude of tweets. The social importance is considered as an indicator of tweets credibility, and  refers to the influence of the microblogger on the social network. The temporal magnitude of microblogs is estimated based on temporal neighbors that present similar query terms. \\
In \cite{chelaru2012}, Chelaru et al. investigated the impact of social features on the effectiveness of video retrieval in Youtube. Social features including likes, dislikes, comments, favorites, etc., are those generated from implicit or explicit interaction with the system. The authors showed that social features are valuable and have a great potential to improve the video retrieval performance when combined to a basic feature like topicality.\\
In social tagging systems, there are also various indicators of relevance representing the users and tags that could be combined in different models.
In a tag recommendation oriented problem, Bel{\'e}m et al. \cite{belem2011} jointly exploited three dimensions of relevance: (i) co-occurrence of terms with tags previously assigned to the object; (ii) terms extracted from multiple textual features, such as title and description; and (iii) relevance metrics such as ``Term Frequency''. In a more recent work \cite{belem2012}, the authors considered both aspects ``Novelty'' and ``Diversity'' to address the problem of tag recommendation. Novel tags are those that are observed very often in the application, and the diversity of a list of recommended tags is estimated by the average semantic distance between each pair of tags in the list. The authors claimed that both aspects are distinct but related concepts. The proposed ranking model combining those metrics has been tested on $3$ popular social media websites: LastFM, Youtube and YahooVideo.
\subsection{On Relevance Aggregation in IR} \label{relevagreg}
Regarding the second line of research that addresses combination of multiple relevance estimates, we identify in the sequel three main strategies.

\textbf{Linear combination approaches.} Researches involving evidence combination from multiple relevance features are often based on linear combination functions \cite{sieg2007,nagmoti2010} due to their simplicity and their relative effectiveness. For instance, in the TREC Microblog tracks \cite{ounis2012}, combination of the different relevance indicators is mostly based on linear functions in the form of products or sums \cite{metzler2011}.
Gerani et \textit{al.} \cite{gerani2012} have proposed a multi-criteria aggregation model allowing to generate a global score that does not necessarily require the comparability of the combinable individual scores. The authors rely on the Alternating Conditional Expectation Algorithm and the BoxCox model to analyze the incomparability problem and perform a score transformation whenever necessary.
More recently, Eickhoff et \textit{al.} \cite{eichkoff2013} introduced a more advanced aggregation mechanism which consists in a statistical framework based on \textit{copulas}. This model addresses the multidimensional relevance assessment, where documents are described by several correlated relevance criteria. The authors compare the copula model to three combination schemes: Sum, Product and weighted linear combination of relevance scores. They argue that these baselines assume independence across all criteria and can be expected to be too naive in some settings where dependence is given \cite{eichkoff2013}. The presented results show that copulas models outperform linear combination mechanisms in two IR tasks namely: opinionated blogs and personalized bookmarks; however, results in another IR application, \textit{i.e.,} retrieval of child-friendly websites, show that linear combination scheme outperforms the copulas models. \\
Another family extensively studied in the literature, is that of ordered weighted averaging functions (Owa) initiated by Yager \cite{yager88}.
In \cite{boughanem2005}, the authors proposed an aggregation scheme that adopt the same idea of the Owa \cite{yager88} operator and uses a weighting method that gives more importance to the terms with high relevance degree to minimize the impact of terms having low scores on the global final evaluation.
%
%
%
%

\textbf{Prioritized aggregation operators.}
These aggregation operators model a priority relationship over the set of criteria. Yager \cite{yager2008} introduced a prioritized scoring operator and a closely related prioritized averaging operator that makes the weights associated to each criterion dependent upon the satisfaction of the higher preferred criterion.
In the IR field, Celia et \textit{al.} \cite{celia2011} proposed a multidimensional representation of relevance through a general prioritized aggregation scheme involving two operators namely, ``And'' and ``Scoring'' \cite{celia2009}. While the Scoring operator is inspired by the prioritized operator proposed in \cite{yager2008}, the ``And'' operator is based on a refinement of the "min" operator. The peculiarity of these operators consists in the fact that if a criterion is not important for the user, its value does not affect the overall performance score, \textit{i.e.,} the weight of a less important criterion should be proportional to the satisfaction degree of more important criteria. The aggregation operators are evaluated in a personalized IR setting. The latter show notable performance improvements when compared to the average operator.

\textbf{Learning to Rank Approaches.}
Instead of proposing carefully designed ranking models based on heuristics principles, a recent emerging thread of research is to apply machine learning algorithms in order to combine multiple relevance features \cite{burges2005,cao2007,hangli2011}.
Given a training set of queries and the associated ground truth containing document labels (relevant, irrelevant), the objective is to optimize the relevance retrieval metrics with the goal of improving the overall search result quality. Optimization is formulated as learning a ranking function in order to minimize a loss function in the training data (eg., number of miss-ordered document pairs).
Each query-document pair is represented by a feature vector consisting of different variables which are functions of the content of the document. 
In the tweet search task, different studies \cite{duan2010,metzler2011,chang2013} propose learning to rank approaches in order to combine several types of twitter features. These work show the benefits of considering machine learning methods in aggregating  relevance criteria. Algorithms like RankSVM and decision tree based learning to rank approaches often show improvements when used in IR tasks \cite{hangli2011}.

However, despite being effective in many IR applications \cite{hangli2011}, these methods tend to offer only limited insight on how to consider importance and interaction between the groups of features that are mapped to different relevance dimensions \cite{eichkoff2013}.

Unlikely, we propose to investigate the combination of general level relevance dimensions using a fuzzy-based aggregation operator. More oriented to the specific problem of relevance aggregation, our method is able to address the property of interaction between dimensions through an integral aggregation operator, namely the Choquet integral, \textit{w.r.t} a fuzzy measure expressing both their individual and joint importance. These properties appear to be appealing from an IR perspective.
In our work \cite{umap2014}, we have successfully proposed a general personalized approach based on Choquet in a contextual suggestion IR task.

\section{Combining relevance estimates With the Choquet Integral}\label{choquet}
\subsection{Choquet-based Relevance Aggregation: Presentation}\label{viewchoq} 
An aggregation operator is a function that maps several inputs from a given interval (\textit{e.g.}, $[0\ldots1]$) to a single output in the same interval. In a typical IR setting, inputs consists in performance scores (RSV\footnote{Retrieval Status Value.}) obtained \textit{w.r.t} each relevance dimension.
In fact, the difficulty in the multidimensional relevance aggregation problem is twofold:
\begin{itemize}
	\item \textit{Estimation of the relevance criteria importance}: identifying which individual criterion and/or subset of criteria need to be enhanced \textit{vs.} weakened regarding the IR task at hand;
	\item \textit{Aggregation and document ranking}: accurately combining the relevance criteria by taking into account their dependency. Based on the documents global scores obtained by combining the partial performance scores \textit{w.r.t} each criterion, it is decided whether a document should be ranked better than another in a ranking.
\end{itemize}
Consider an IR scenario where $ \mathcal{D}=\{d_{1}, d_{2}, \ldots, d_{M} \}$ is the set of documents, $\mathcal{C}$ $=$ $\{c_{1}, c_{2}, \ldots, c_{N} \}$ is the set of relevance criteria and $q$ a given query.
Let \textit{RSV}$_{c_{i}}$($q$,$d_{j}$) be the performance scores of document $d_{j} \in \mathcal{D}$, obtained \textit{w.r.t} the relevance criterion $c_{i} \in \mathcal{C}$. The task of combining the performance scores of  $d_{j}$ \textit{w.r.t} all $c_{i} \in \mathcal{C}$ is called \textit{aggregation}.  
Formally, the general aggregation function, denoted by $\mathcal{F}: \mathbb{R}^{N} \longrightarrow \mathbb{R}$, that computes the global score of document $d_{j}$ in response to query $q$, is defined as follows.
\begin{dmath}\label{eqchoq}
\textit{RSV}_{c_{1}}(q,d_{j})  \times \ldots \times \textit{RSV}_{c_{N}}(q,d_{j})) \longrightarrow  \mathcal{F}(\textit{RSV}_{c_{1}}(q,d_{j})\\, \ldots,
\textit{RSV}_{c_{N}}(q,d_{j}))
\end{dmath}
Where \textit{RSV}$_{c_{i}}$($q,d_{j}$) is the performance score of $d_{j}$ \textit{w.r.t} criterion $c_{i}$.
As we rely on the Choquet operator, to each subset of criteria is associated a fuzzy measure, also called \textit{capacity}, that reflects its importance. \\

\begin{definition}
Let $I_{\mathcal{C}}$ be the set of all possible subsets of criteria from $\mathcal{C}$.
A fuzzy measure is a normalized and a monotone function $\mu$ from $I_{\mathcal{C}}$ to $[0 \ldots 1]$ such that:\\
 $\forall$ $I_{C_{1}}$, $I_{C_{2}}$ $\in$ $I_{\mathcal{C}} $, if $(I_{C_{1}}\subseteq I_{C_{2}})$ then $\mu(I_{C_{1}})$ $\leq$ $\mu(I_{C_{2}})$, with $\mu (I_{{\varnothing}}) $ = $0$ and $\mu (I_{\mathcal{C}})$ = $1$. \\
\textnormal{For the sake of notational simplicity, $\mu(I_{C_{i}})$  will be denoted by $\mu_{C_{i}}$. The value of $\mu_{C_{1}}$ can be interpreted as the importance degree of the interaction between the criteria involved in the subset $C_{1}$.}
\end{definition}
The Choquet integral based-relevance aggregation function built on such a fuzzy measure is defined as follows.
\begin{definition}
\begin{dmath}\label{eqchoq2}
	Ch_{\mu}(\textit{RSV}_{c_{1}}(q_{r},d_{j}),\ldots, \textit{RSV}_{c_{N}}(q_{r},d_{j})) \\
= \sum_{i=1}^{N}\mu_{c_{i},..., c_{N}}.(\textit{rsv}_{(i)j}-\textit{rsv}_{(i-1)j})
\end{dmath}
Where $Ch_{\mu}$ is the Choquet aggregation function, $rsv_{(i)j}$ is the $i-th$ element of the permutation of $RSV(q_r,d_j)$ on criterion $c_{i}$, such that ($0 \leq rsv_{(1)j} \leq ...\leq rsv_{(N)j}$).
\end{definition}
The value of $\mu_{I_{C_{i}}}$ can be interpreted as the importance degree of the interaction between the criteria involved in subset $C_{i}$. Note that if $\mu$ is an additive measure, the Choquet integral corresponds to the weighted mean. Otherwise, it requires fewer than ($2^N-1$) capacity measures in the case where the fuzzy measure is $k-$order additive, \textit{i.e.,} $\mu_{A}=0$ for all criteria subsets $A \subseteq \mathcal{C}$ with $\left| A \right| > k$.

As previously stated, the Choquet operator exhibits a number of properties that appear to be appealing from an IR point of view. From a theoretical perspective, since it is built on the concept of fuzzy measures, it allows modeling flexible interactions and considering complex dependencies among criteria \cite{grabisch2000}.
To facilitate the task of interpreting the Choquet integral behavior, we exploit two parameters namely, the ``importance indice'' and the ``interaction indice'' \cite{grabisch2000} that offer readable interpretations and qualitative understanding of the resulting aggregation model. For more details on the interaction and importance indices, see our extended work in \cite{moulahietalj2014}.

\begin{definition}\label{defshapley} \textit{Importance index: }
Let $\mu_{c_{i}}$ be the weight of relevance criterion $c_{i}$ and $\mu_{Cr \cup c_{i}}$ its marginal contribution to each subset $Cr \in \mathcal{C}$ of other criteria. The importance index \cite{shapely53} of $c_{i}$ \textit{w.r.t} a fuzzy measure $\mu$ is then defined as the mean of all these contributions:
\begin{dmath}\label{shap}
 \phi_{\mu} (c_{i}) = \sum_{Cr \subseteq \mathcal{C} \setminus \{c_{i}\}} \frac{(N - |Cr| - 1)!.|Cr|! }{N!} [\mu_{Cr}.\mu_{(Cr \bigcup c_{i})}]
\end{dmath}
$\phi_{\mu} (c_{i})$ measures the average contribution that criterion $(c_{i})$ brings to all the possible combinations of criteria.
\end{definition}
By introducing this indice, the overall importance of criterion $c_{i}$ is no longer solely determined by its weight $\mu_{c_{i}}$ but also by its contribution to each subset of other criteria.

\begin{definition}\label{definteract} \textit{Interaction index: }
Let $(\Delta_{c_{i}c_{j}}\mu_{Cr})$, with $Cr$ = $\mathcal{C} \setminus \{c_{i},c_{j}\} $, be the difference between the marginal contribution of criterion $c_{j}$ to every combination of criteria that contains criterion $c_{i}$, and a combination from which criterion $c_{i}$ is excluded:
\begin{dmath}\label{int1}
(\Delta_{c_{i}c_{j}}\mu_{Cr})  =  [\mu_{(\{c_{i}c_{j} \} \cup Cr)} - \mu_{(c_{i} \cup Cr)}] - [ \mu_{(c_{i}\cup Cr)} - \mu_{Cr} ] \\
\end{dmath}

\textnormal{This expression is defined to appraise the strength among two criteria $c_{i}$ and $c_{j}$. When this latter expression is positive (\textit{resp.} negative) for any $Cr \in \mathcal{C} \setminus \{c_{i},c_{j}\}$, we say that both criteria
$c_{i}$ and $c_{j}$ positively (\textit{resp.} negatively) interact (\textit{i.e.,} the contribution of criterion $c_{j}$ is higher with the presence of criterion $c_{i}$).\\
The interaction index among two measures is thus defined as follows}:
\begin{dmath}\label{int2}
I_{\mu}(c_{i},c_{j})  =  \sum_{Cr \subseteq \mathcal{C} \setminus \{c_{i},c_{j} \}  } \frac{(N - |Cr| - 2)!.|Cr|!}{(N-1)!} (\Delta_{c_{i}c_{j}}\mu_{Cr})
\end{dmath}
\end{definition}
The interaction value, which falls into the interval $[-1..1]$, is zero when both criteria are independent and it is positive (\textit{resp.} negative) whenever the interaction between them is positive (\textit{resp.} negative).

\subsection{Training the fuzzy measures within an IR task}\label{learnalgo}
	
\begin{figure*}[htbp]
  \begin{center}
    \includegraphics[width=0.9\textwidth]{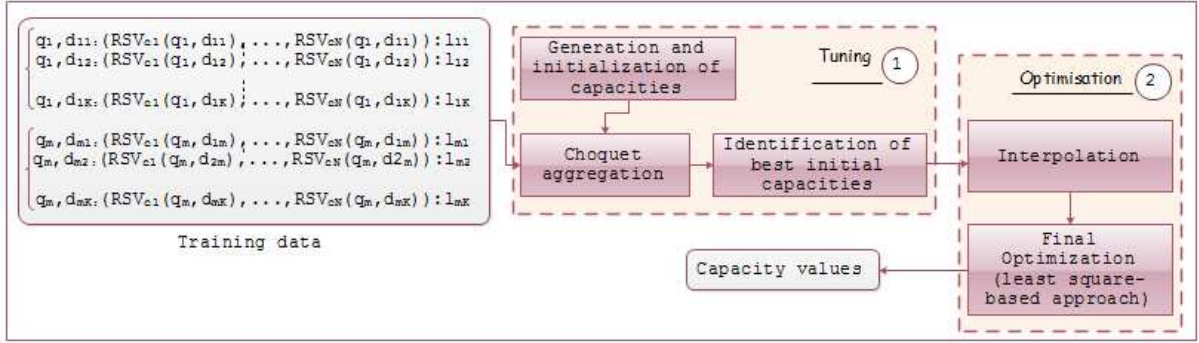}
  \end{center}
  \vspace{-15pt}
  \caption{General Paradigm of Training the Fuzzy Measures (Capacities) for all Criteria and Subsets of Criteria}\label{capacityidentif}
	\end{figure*}

The objective of the training step here is to optimize the fuzzy measures \textit{w.r.t} a target IR measure (e.g. $P@X$) by identifying the values of the Choquet capacities to be used in the aggregation process.
Figure \ref{capacityidentif} illustrates the adopted methodology. As in machine learning algorithms, the typical training data required for learning the Choquet fuzzy measures includes a set of training queries $q_{k}$ and for each query, a list of ranked documents $d_{kj}$ represented by pre-computed vectors containing performance scores $RSV_{c_{i}}(q_{k},d_{kj})$. \\ $RSV_{c_{i}}(q_{k},d_{kj})$ denotes the score of $d_{kj}$ \textit{w.r.t} criterion $c_{i}$ in response to query $q_{k}$. Each document is annotated with a rank label $l_{kj}$ (\textit{e.g.,} relevant or irrelevant). As presented in Figure \ref{capacityidentif}, the training involves two main steps, described in what follows.

\textit{\textbf{Tuning:}} the objective of this step is to generate and tune the capacity values in order to identify the best initial capacity values. In a first stage, we start by generating and initializing a set of capacity combinations, where each capacity combination $\mu^{(.)}$ represents the set of capacity values assigned to each criterion and subset of criteria. In the case of $N$ relevance criteria, each capacity combination includes ($2^N-1$) capacity values. For instance, if we consider three relevance criteria, a capacity combination involves the values ($\{\mu_{c_{1}};\mu_{c_{2}}; \mu_{c_{3}};\mu_{c_{1},c_{2}};\mu_{c_{1},c_{3}}$; $\mu_{c_{2},c_{3}}\}$).
	Those values fall into $[0\ldots1]$ and are tuned for all criteria, such that their sum is equal to $1$. The value are computed with a step equal to $0.1$. In the case of three criteria, we fix the capacity value of a criterion (\textit{e.g.,} $0.1$) and we tune the remaining values (\textit{e.g.,} from $0.8$ to $0.1$ for the first; and from $0.1$ to $0.8$ for the second). the process is repeated for each criterion. The tuning is conceivable since there is generally a few relevance dimensions \cite{saracevic2007III}. However, when the number of criteria is strictly higher than $3$, we can avoid the tuning complexity by relying on sub-families of capacities namely $2$-additive measures \cite{grabisch2000}, requiring less coefficients to be defined and assuming that there is no initial interaction among subsets of more than $2$ criteria. In a second stage, we used the Choquet integral operator to aggregate the different relevance criteria estimates of the performance scores in the training data \textit{w.r.t} to each generated capacity combination $\mu^{(.)}$. Then, we select the best capacity combination $\mu^{(*)}$ giving the best results \textit{w.r.t} a target IR measure (\textit{e.g.}, $P@X$).
	
\textit{\textbf{Optimisation:}} the objective of this step is to optimize the initial capacity combination $\mu^{(*)}$ obtained in the previous step. We start by $\mu^{(*)}$, we pull the top $K$ documents returned by each training query $q_{k}$ and we interpolate the documents scores to boil down the non relevant ones. Interpolation is based on performance scores of the training documents based on $\mu^{(*)}$. The underlying objective is to affect relevant documents higher scores than non relevant ones.
Finally, we proceed to the application of the Least-squares based optimization. Obtained capacity values are used within each relevance criterion and subset of relevance criteria.

\section{Experimental design}\label{experiments}
Experimental evaluation is based on two IR evaluation frameworks, namely the tweet search task \cite{ounis2011} and the personalized social bookmarking IR task \cite{vallet2012}. This section describes these tasks, the used data, baselines and metrics.

\subsection{Tasks}
\subsubsection{Tweet Search Task.}\label{tweetsearch}
We exploit the dataset and topics of the TREC $2011$ and $2012$ Microblog tracks. The datasets include more than $16$ million tweets and more than $5$ million users \cite{ounis2011}. TREC Microblog $2011$ track includes (49) topics, used for training the capacity values and TREC Microblog $2012$ includes ($60$) topics used for testing.  We make use of three important and widely used relevance criteria \cite{nagmoti2010}: \textit{topicality}, \textit{recency} and \textit{authority}. To deal with the topicality relevance criterion, we propose to use the Okapi BM25 ranking, then, we introduce the authority and the recency relevance dimensions as query independent measures, as done in \cite{nagmoti2010}. The analysis of the criteria importance using the importance indice (Cf., Section \ref{viewchoq}) \cite{grabisch2000} reveals a high importance of topicality with a value of $0.631$. The recency relevance criterion is also given a quite high importance (of about $0.25$) compared to the authority relevance dimension ($0.12$). This is not surprising as far as a user usually seeks for topically relevant documents rather than those which are authoritative or even more recent.
\begin{figure}[htbp]
  \begin{center}
    \includegraphics[width=6.8cm,height=4.8cm]{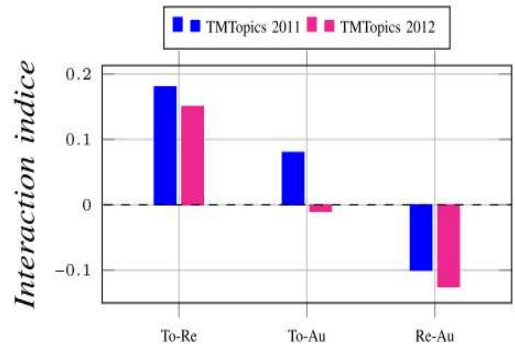}
  \end{center}
  \caption{Criteria Interaction Indice.}\label{corrinteract}
	\end{figure}

Figure \ref{corrinteract} shows the values of the interaction indice between topicality, recency and authority, denoted respectively by \textit{To, Re} and \textit{Au} within both TREC Microblog $2011$ and $2012$ track topics (referred respectively as ``TMTopics $2011$'' and ``TMTopics $2012$'').
From this Figure, we can see that the authority criterion does not bring any contribution when it is combined with topical relevance criteria.
Moreover, we notice a positive interaction between the \textit{topicality} and \textit{recency} relevance criteria. This explains the higher contribution of these two criteria on the overall global scoring when they are present together, and allows us to evaluate our approach under the presence of interdependent relevance criteria.

\begin{figure*}[!ht]
\subfloat[Capacities values effectiveness within the tweet search task.
	The $x$-axis represents some of the $21$ trained capacities combinations.]{\includegraphics[width=1\textwidth]{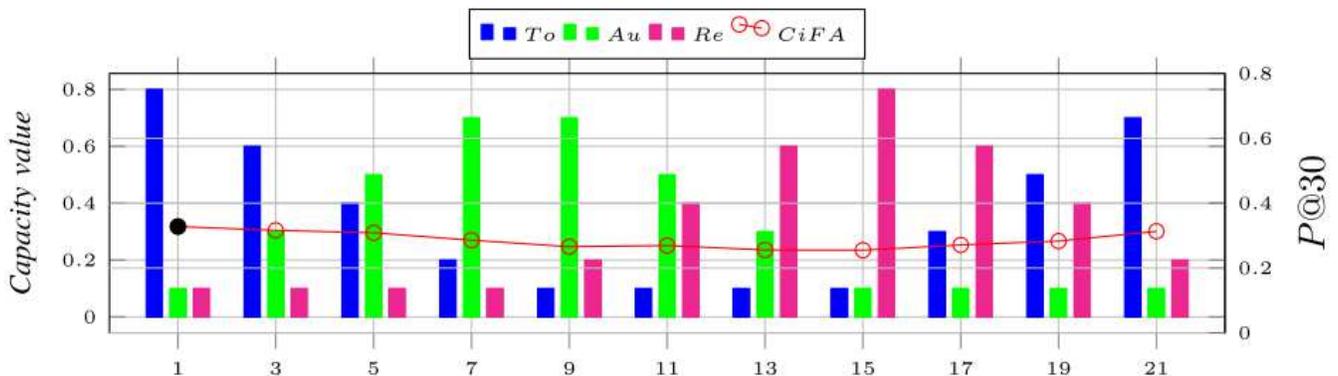}\label{figfuzzy}}
	
\subfloat[Capacities values effectiveness within the social personalized task.]{\includegraphics[width=1\textwidth]{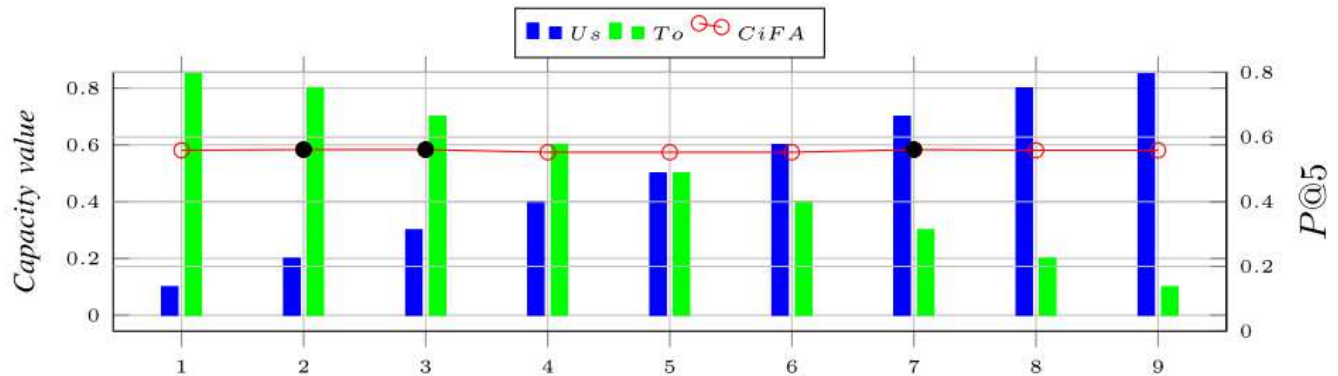}\label{fuzzytrainperso}}
\caption{Capacities tuning within both datasets.}
\label{trainingfuzz}
\end{figure*}

\subsubsection{Personalized social bookmarking task.}\label{persosearch}
With the advent of the Web 2.0, social tagging systems, such as \textit{e.g.,} \textit{Del.icio.us}\footnote{http://www.delicious.com} and \textit{Flickr}\footnote{http://www.flickr.com}, have exponentially grown both in terms of users and contents. We exploit a compiled collection of about $33$k delicious bookmarks and $12$k tags.\footnote{http://ir.ii.uam.es/~david/webdivers/} The dataset contains evaluation information from $35$ users over $177$ search topics \cite{vallet2012} according to two relevance dimensions namely, the topical relevance ($To$) of bookmarks given a topic and its personal or user relevance ($Us$) given a user. In our evaluation protocol, we use $75\%$ of the topics for learning the capacity values and we exploit the remaining topics for the testing phase.

We compute the importance indices and we found that both of them have quite similar importance with indices of about $0.48$ for the topical relevance and $0.51$ for the user one. The interaction value between the considered relevance criteria ($To-Us$) is about $0.028$, which is quite low to argue that they are really independent.

\subsection{Baseline runs and metrics}
Firstly, we compare our approach to classical aggregation operators such as the linear combination scheme (\textsc{Lcs}) and the \textsc{Owa} \cite{yager88}, as well as the two competing prioritized operators namely, \textsc{And} and \textsc{Scoring} \cite{celia2011}. Then, we compare it to state-of-the-art learning to rank algorithms namely \textsc{RankSVM} \cite{joachims2006}, \textsc{Random Forest} \cite{breiman2001} and $\lambda-$\textsc{MART}. With respect to each task specificities and guidelines, we used: 1) for the tweet search task, the $P@5$, $P@10$, $P@20$, $P@30$, $P@100$ and MAP measures. We emphasize that the official measure of the track is P@30;  2) for the personalized bookmarking task, given that only the top $5$ relevant results are provided with each topic, we make use of the $P@5$ measure as recommended in \cite{vallet2012}. In these experiments, significance testing is based on the t-student statistic for both datasets.

\begin{table*}[htbp]
\centering
\begin{tabular}{lrrrrrlcc}
\toprule
   &\multicolumn{5}{c}{\textbf{Precision}}&  & \\ \cmidrule(r){2-6}
  \textbf{\textsc{Operator}} &  \textbf{P@5}  & \textbf{P@10} & \textbf{P@20} & \textbf{P@30} & \textbf{P@100} & \textbf{MAP} & \multicolumn{2}{c}{\textbf{\% change}} \\

\midrule
 \textbf{\textsc{LCS}} & $0.1965$& $0.1860$ & $0.1833$   & $0.1854$ & $0.1309$ & $0.0928$ & $\textbf{+20,73\%}$ & $\S$  \\\cmidrule{1-9}
 \textbf{\textsc{OWA}}   & $0.1828$  & $0.1879$ & $0.1767$ & $0.1764$ & $0.1248$ & $0.0882$  & $\textbf{+23,73\%}$ & $\star$\\\cmidrule{1-9}

\cmidrule{1-9}
\textbf{\textsc{AND}} & $0.1828$ & $0.1793$ & $0.1767$ & $0.1764$   & $0.1233$  & $0.0882$ & $\textbf{+23,73\%}$ &  $\star$ \\
\textbf{\textsc{SCORING}}& $0.2000$ & $0.2018$  &  $0.1982$  & $0.1977$& $0.1475$& $0.1091$  & $\textbf{+14,52\%}$ & $\star$  \\\cmidrule{1-9}
\textbf{\textsc{RANKSVM-L}} & $0.2207$ & $0.2276$  &  $0.2207$ &$0.2213$ & $0.1586$& $0.1213$  & $\textbf{+4,32\%}$ & $\star$  \\

\textbf{\textsc{RANKSVM-RBF}} & $0.0966$ & $0.0828$  &  $0.0733$ &$0.0856$ & $0.0993$& $0.0665$  & $\textbf{+62,99\%}$ & $\star$  \\

\textbf{\textsc{RF}} & $0.1000$ & $0.0810$  &  $0.0681$ &$0.0687$ & $0.0628$& $0.0464$  & $\textbf{+70,68\%}$ & $\star$  \\

\textbf{$\lambda-$\textsc{MART}} & $0.2931$ & $0.2276$  &  $0.2092$ &$0.2043$ & $0.1856$& $0.1321$  & $\textbf{+11,67\%}$ & $\star$  \\

\cmidrule{1-9}

 \multirow{2}{*}{\textbf{\textsc{CiFA}}} & $\textbf{0.2379}$ & $\textbf{0.2362}$  &  $\textbf{0.2422}$ & $\textbf{0.2313}$ & $\textbf{0.1614}$& $\textbf{0.1295}$  & \multicolumn{2}{c} {$-$}\\\cmidrule{2-7}
	   & $\textbf{+7.22\%}$ & $\textbf{+3.64\%}$  &  $\textbf{+8.87\%}$ & $\textbf{+4.32\%}$ & $\textbf{+1.73\%}$& $\textbf{+6.33\%}$  &   \multicolumn{2}{c} {$\star$} \\
	\bottomrule
\end{tabular}
\centering
\vspace{0.2cm}
\caption{Comparative evaluation of retrieval effectiveness. $\%$ change indicates the \textsc{CiFA} improvements in terms of $P@30$. The symbols $\S$ and $\star$ denote the student test significance: $"\S"$: $0.05< t \leqslant 0.1$; $"\star"$: $t \leqslant 0.01$. The last row shows the \textsc{CiFA} improvement in terms of $P@X$ and MAP with the best baseline (\textit{i.e.,} \textsc{RankSVM}).}
\label{tablemesure}
\end{table*}

\begin{table*}[!htbp]
\centering
\begin{tabular}{lcccccc}
\toprule
  & \textbf{\textsc{LCS}} & \textbf{\textsc{OWA}} & \textbf{\textsc{AND}} &  \textbf{\textsc{SCORING}}& \textbf{\textsc{RankSVM}} & \textbf{\textsc{CiFA}} \\
\midrule
\textbf{P@5} &   $\textbf{0.6310}$  & $\textbf{0.6310}$ & $0.6286$ &   $\textbf{0.6310}$     &  $0.6286$    &  $\textbf{0.6310}$ \\
\midrule

\multirow{2}{*}{\textbf{\% change}}  &  $0 \%$   &   $0 \%$ & $\textbf{+0,003\%}$ &   $0 \%$      & $\textbf{+0,003\%}$     & \multirow{2}{*}{$-$}  \\ \cmidrule{2-6}

	   & $\star\star\star$ & $\star\star\star$ & $\star\star\star$& $\star\star\star$ &  $\star\star\star$ &  \\

	\bottomrule
\end{tabular}
\centering
\vspace{0.2cm}
\caption{Comparative evaluation of retrieval effectiveness within the personalized social bookmarking task. The symbol $\star$ denotes the student test significance: $"\star\star\star"$: $t \leqslant 0.01$.}
\label{tpersscores}
\vspace{-0.3cm}
\end{table*}

\section{Results and Discussion}\label{evalresult}
 \subsection{Training the fuzzy measures}\label{traincap}

To identify the Choquet capacity values, we adopt the methodology illustrated in Figure \ref{capacityidentif} (Cf., Section \ref{learnalgo}). The value of the top retrieved documents $K$ is set to $100$. For the tweet search task, we instantiate $Q_{learn}$ and $qrels$ by the TREC Microblog $2011$ topics and the corresponding relevance assessments respectively and use the official measure $P@30$ for the tuning. For the personalized social bookmarking task, we used random $75\%$ of the topics for the training phase and use the $P@5$ measure for the tuning. The trained combination capacities in this learning phase, will be denoted by $\mu^{(i)}$. The initial capacity value of each criterion is obtained with a step of $0.1$ such that the sum of the three capacities is $1$, leading to $21$ and $9$ capacity values respectively for the tweet search task and the personalized bookmarking task. Then, we select the combination ($\mu^{(*)}$) achieving the highest average value of the precision measure. Figure (3a) and (3b) give an overview of the training results for the tweet search dataset and the social bookmarking dataset respectively; the optimal combination $\mu^{(*)}$ is highlighted in black.

From Figure (3a) we can see that, as expected from the importance and interaction analysis presented in Sections \ref{tweetsearch} and \ref{persosearch}, the tweet search task is more sensitive to the topical criterion $To$ than the authority $Au$ and recency criteria $Re$, including values of about $0.8$ for $\mu_{To}$ and $0.1$ for $\mu_{Au}$ and $\mu_{Re}$, respectively.
After the application of the Least squares based optimization method, we obtained $\mu^{(**)}$ which is composed by: ($\mu_{To} = 0.705$, $\mu_{Re}= 0.215$, $\mu_{Au}= 0.025$, $\mu_{\{To,Re\}}=0.973$, $\mu_{\{To,Au\}}= -0.14$, $\mu_{\{Re,Au\}}= -0.25$). We notice that the capacity values obtained for $\mu^{(**)}$ on the subsets $\{To,Au\}$ and $\{Re,Au\}$ are negative ones. This suggests that the authority criterion does not appear to be a good factor when combined with topicality or recency, and this explains the negative capacities assigned to $\mu_{\{To,Au\}}$ and $\mu_{\{Re,Au\}}$, which corroborates previous results \cite{nagmoti2010}.

From Figure (3b) we observe that the $P@5$ values \textit{w.r.t} to the different capacity combination values are quite close.
It comes somewhat unexpected, that the $P@5$ is either high when the user relevance ($Us$) is quite more important $(\mu^{(7)})$ or even when the topical relevance ($To$) is highly valued $(\mu^{(2)}, \mu^{(3)})$. As the $P@5$ is reached for $(\mu^{(7)})$ and the latter yields more importance to the user relevance, we select it as an initial capacity for identifying the fuzzy measures with the Least squares based-optimization method. The capacity values returned are quite similar to those obtained by the importance indice in Section \ref{persosearch}, giving a capacity ($\mu_{Us}$) of about $0.5$ to the user relevance, $0.47$ for the topical one ($\mu_{To}$) and $0.028$ for the subset of both criteria ($\mu_{\{Us,To\}}$).

\subsection{Measuring the Retrieval Effectiveness}\label{effectiveness}
This subsection presents the results of the testing phase. For this aim, we use the $60$ topics from the TREC Microblog $2012$ track, and the $25\%$ remaining topics of the personalized bookmarking task. The baseline parameters have been tuned using the same learning datasets used for the choquet based operator. For the learning to rank method, we used the open source code of \textsc{RankSVM} from \cite{joachims2006}.

\subsubsection{Relevance Estimation Within the Tweet Search Task}\label{effectivenesstweet}

Table \ref{tablemesure} reports the retrieval performance obtained by our choquet based aggregation operator, denoted \textsc{CiFA} in the remainder, in comparison with the baselines.
Note that we ran a series of experiments to optimize the learning to rank methods parameters. Those experiments are run with five cross validation within the same learning set used to find the Choquet capacities values. For \textsc{RANKSVM} we tested the linear (\textsc{RANKSVM-L} model and the Radial Basis Function (\textsc{RANKSVM-RBF)} kernel. Then, we have tried different values of the parameter $C$ ($10^{-5}..10^{2}$) for the former, and various pairs of ($C$, $\sigma$) for the latter.  Through cross validation, we found that $C=0.00005$ gives the best performance for the linear \textsc{RANKSVM-L} and found that the pair ($8.88$,$7.77$) is the best setting for the RBF kernel. For the tree-based learning to rank methods, we found that they are quite insensitive to parametrization. Using cross validation, we found that a number of trees of about $1000$, a learning rate equal to $0.1$ and a number of leaves equal to $2$ gives the best performance for $\lambda-$\textsc{MART}.

From Table \ref{tablemesure}, we can see that \textsc{CiFA} overpasses all the baselines. We notice that the performance improvements are more important for the classical aggregation operators; they reach $23,73\%$ compared to the \textsc{OWA} operator and $20,73\%$ compared to the \textsc{LCS} operator. For the \textsc{Scoring} operator, the significant improvement is less important. As we considered the prioritization scenario $Sc_{1}$: $\{topicality\} \succ \{recency\} \succ \{authority\}$, giving the best $P@30$ average over the other possible prioritization scenarios, we can conclude that the obtained difference of performance, in favor of \textsc{CiFA}, is explained by the consideration of the interactions existing among the set of criteria, that we involved by means of the fuzzy measures. Thus, the global scores can no longer be biased by dependent relevance criteria or overestimated by those highly scored than the other ones. For the \textsc{And} operator, the improvement difference is sharply better. The obtained results are likely due to the fact that it is mainly based on the \textsc{Min} operator, which could penalize tweets highly satisfied by the least important criteria. Roughly speaking, if there are many tweets highly scored \textit{w.r.t} the authority criterion (which is likely the case), its overall satisfaction degrees would be biased by this relevance criterion.
Last but not least, we can see that \textsc{CiFA} significantly outperforms the linear \textsc{RankSVM} which represents here the best baseline. The improvement varies between $4.32\%$ to $8.87\%$, and it is quite high for the $P@5$, $P@10$ and $P@20$, enhancing thus the quality of the first retrieved tweets. Improvements for the other learning to rank algorithms is more important. They are higher for \textsc{RANKSVM-RBF} and \textsc{RF} but less important for $\lambda-$\textsc{MART}, with an improvement of about $11,67\%$.

\subsubsection{Relevance estimation within the social personalized IR task.}\label{perseffect}
As we can see from Table \ref{tpersscores}, the performance of \textsc{CiFA} as well as the baselines are quite close. Apart from the evaluation measure used, this is likely due to the number of criteria involved as well as the slight independence existing between them, as shown through the interaction analysis (Cf. Subsection \ref{persosearch}). Therefore, the behavior of the Choquet integral is quite similar to that of a linear combination scheme which explains the results reported in Table \ref{tpersscores}. Interestingly, these results also highlight the fact that the Choquet operator performance is stable despite criteria independence and leads to comparable results than the baseline operator results at least when the number of criteria is quite small.

 \section{Conclusion and Future Work}\label{conclusion}
We presented a novel general multi-criteria framework for multidimensional relevance aggregation. Our approach relies on a fuzzy integral method based on the well studied and theoretically justified Choquet mathematical operator. The proposed operator supports the observation that relevance dimensions, measurable through criteria, may interact and have different weights (importance) according to the task at hand. Besides, the approach  allows analyzing the resulting model criteria behavior with a readable interpretation through the Shapley and interaction indices.
A set of experiments were conducted on TREC Microblog datasets as well as a personalized social bookmarking corpus and showed that our operator significantly outperforms both supervised and unsupervised aggregation methods. The main limitation of this work concerns the computational complexity of the learning algorithm in the case where the number of relevance dimensions is high. This specific issue can be addressed by considering other properties of the Choquet integral to be drawn within the document rankings. \\
In future, we plan to investigate how to integrate the user or class of users (eg. children, students) as a dependent variable within the framework in order to design personalized aggregated rankings. It is possible to achieve this by learning a user-based fuzzy measure and assume independence between user classes.
\vspace{-0.1cm}
\bibliographystyle{abbrv}
\bibliography{these}

\end{document}